\documentclass[11pt]{article}
\pdfoutput=1

\newif\ifanon
\ifdefined\anonmode \anontrue \fi
\newif\ifreview
\ifdefined\reviewmode \reviewtrue \fi
\ifanon \reviewtrue \fi

\usepackage[T1]{fontenc}
\usepackage{amsthm}
\usepackage{newtxtext,newtxmath}
\usepackage[margin=2.5cm]{geometry}
\ifreview\usepackage[onehalfspacing]{setspace}\else\usepackage{setspace}\fi
\usepackage[switch]{lineno}
\theoremstyle{definition}
\newtheorem{definition}{Definition}[section]
\usepackage{booktabs}
\usepackage{tabularx}
\usepackage{listings}
\usepackage[obeyFinal]{todonotes}
\usepackage{tikz}
\usepackage{tcolorbox}
\newtcolorbox{rqanswer}[1]{colback=gray!8, colframe=gray!60, boxrule=0.5pt, arc=2pt,
  left=4pt, right=4pt, top=3pt, bottom=3pt, fonttitle=\bfseries\small, title={Answer to #1}}
\usetikzlibrary{arrows.meta}
\usepackage[style=apa,backend=biber]{biblatex}
\let\cite\parencite
\usepackage[hidelinks]{hyperref}
\usepackage{cleveref}

\newcommand{\harness}{\ifanon AnonHarness\else agentknit\fi}
\newcommand{\harnessurl}{\ifanon name and link anonymized for review\else\url{https://github.com/monperrus/agentknit}\fi}
\newcommand{\counter}{\ifanon\texttt{anon-trajectories}\else\texttt{reproducible-trajectories}\fi}
\newcommand{\selfcite}[1]{\ifanon\parencite{anonymous2026}\else\parencite{#1}\fi}
\newcommand{\selftextcite}[1]{\ifanon\textcite{anonymous2026}\else\textcite{#1}\fi}

\title{Knowledge as Code}
\ifanon
\author{}
\else
\author{Martin Monperrus\\KTH Royal Institute of Technology, Stockholm, Sweden\\\href{mailto:monperrus@kth.se}{monperrus@kth.se}}
\fi
\date{}

\begin{document}

\maketitle

\begin{abstract}
For millennia, knowledge has been carried by language, and large language models still put language at the center.
Yet the coding agents built on these models mostly write code.
This paper is a contribution to epistemology, which asks what kind of knowledge code is and how it is obtained.
We define knowledge as code as knowledge written in a programming language or in a machine-parsable data format, whose validity can be checked by executing it.
We define thirteen epistemic properties of knowledge media and show that code is the first medium to have all of them.
We distinguish two ways in which knowledge enters code, encoding and learning by execution, where code is both the experiment and the record of its result.
Agents make learning by execution observable, since their trajectories record every probe, observation and line of code they keep.
Code is thus not only a tool of knowledge work but a knowledge medium in its own right, the first that can check what it holds.
\end{abstract}

\ifreview\noindent\textbf{Keywords:} epistemology, knowledge as code, knowledge media, coding agents, learning by execution, philosophy of computing\fi

\ifreview\linenumbers\fi

\section{Introduction}
\label{sec:intro}

For millennia, knowledge has been carried by language, first in living memory and then in writing \cite{ong1982orality}.
Large language models continue this tradition, and their name says it: they learn from language and answer in language, and the facts they hold can be queried in language \cite{petroni2019language}.

Philosophy has studied knowledge carried by speech and by texts \cite{ong1982orality,goody1963consequences}, and machine learning research studies the knowledge held in the weights of language models \cite{petroni2019language}.
\textbf{Neither has an account of code as a carrier of knowledge}, although a growing part of what is known about the digital world is written in code.

When an agent queries an API, reads the response, and writes a parser for it, something has been learned about the world and kept.
The parser is knowledge of the API, justified by the response that the agent observed, and checked every time the parser runs, which is not true of the same fact written in prose or held in weights.

We propose knowledge as code, defined as knowledge written in a programming language or in a machine-parsable data format, whose validity can be automatically checked.
We define thirteen epistemic properties of knowledge media: persistence, transmissibility, precision, expressiveness, inspectability, editability, provenance, composability, executability, verifiability, reproducibility, generativity and scalability.
We apply them to oral tradition, written language, LLM weights and code.
Code meets all thirteen; code is the first knowledge medium that keeps what writing brought millennia ago and adds to it.
When knowledge is code, knowledge can act, be checked by a machine, be checked again later, and produce new knowledge.

We distinguish two ways in which knowledge enters code, encoding, which moves existing knowledge into code, and learning by execution, which obtains new knowledge by executing code against the world.
We describe the executable knowledge lifecycle that produces, checks and revises the latter.

We measure these concepts on the work of coding agents.
Agents make knowledge as code observable, since their trajectories record every probe, every observation and every line of persisted code.
Our corpus holds 29,260 trajectories and our instruments are fully automated.
We measure how much of what agents write is knowledge as code, how much of the code they keep was learned by execution, how much of it is knowledge verification, and whether what they learned still holds months later.

This paper is a contribution to epistemology, which asks what kind of knowledge code is and how it is obtained.
No prior work gives an epistemic account of code, one that compares it with other media by their properties and explains how knowledge enters it by execution.

This paper makes the following contributions.
\begin{itemize}
\item Code is shown to be the first knowledge medium that has all thirteen epistemic properties (\cref{sec:properties,sec:comparison}).
\item In learning by execution, code is both the experimental means and the experimental outcome (\cref{sec:kac}).
\item Empirical statistics about knowledge-as-code. Four lines out of five that coding agents write are knowledge as code, and one line of code in four is a probe.
\end{itemize}

\section{Background}
\label{sec:background}

\subsection{Code}
\label{sec:code-data}

Code is first text written in a programming language.
A programming language is a formal language with a syntax that a machine parses and a semantics that a machine carries out, through an interpreter or a compiler.
A program therefore has two readers.
A human reader recovers what the program means, and Knuth asks programmers to write for this reader: ``let us concentrate rather on explaining to human beings what we want a computer to do'' \cite{knuth1984literate}.
A machine reader carries the program out.
Text in natural language has only the first reader.

Structured data is text in a language that has a grammar but no execution semantics of its own.
For example, JSON \cite{bray2017json} and TOML \cite{prestonwerner2021toml} are such languages, used for data exchange and configuration files.
A parser accepts or rejects a document in such a language without ambiguity.
Parsing is thus a mechanical verification, since a malformed document is detected before any use.
A program then consumes the document, and structured data acts through the code that reads it.

We group programs and structured data under the name code because they share what matters for this paper, namely that a machine reads them and acts on them.

\subsection{Epistemology}
\label{sec:epistemology}

Epistemology is the study of knowledge.
The term comes from the Greek words ``episteme'' and ``logos'', which can be translated as knowledge and account \cite{steup2024epistemology}.
Epistemology gives the vocabulary to ask what kind of knowledge code holds and how that knowledge is obtained.

This paper uses the following core terms.
\begin{description}
\item[Knowledge] In the classical analysis, ``$S$ knows that $p$ if and only if $p$ is true and $S$ justifiably believes that $p$'' \cite{steup2024epistemology}.
Popper's objective knowledge has no believer.
It is made of the claims and arguments kept in ``journals, books, and libraries'' \cite[p.~107]{popper1972objective}, and it ``is knowledge without a knower'' \cite[p.~109]{popper1972objective}.
This paper uses knowledge in this objective sense, as a claim, its justification and its truth.
\item[Claim] What is held to be true, here a statement about the world.
\item[Justification] The reasons that support a claim; in empirical knowledge, the observations on which the claim rests.
\item[Observation] What an experiment or a measurement shows about the world at the moment it is made.
\item[Verification] The act of checking whether a claim is true, by comparing it with the world.
\end{description}

\section{Epistemic Properties of Knowledge Media}
\label{sec:properties}

We first define what a knowledge medium is.
We then define thirteen epistemic properties, each with a test question that decides whether a medium has it.
\Cref{sec:comparison} applies the properties to four media and backs each verdict with a citation or a measurement.

\subsection{Knowledge Media}

Popper distinguishes three worlds: the world of physical objects, the world of mental states, and the world of ``objective contents of thought'' \cite[p.~106]{popper1972objective}.
Objective knowledge belongs to the third world.
We call \emph{knowledge medium} a material form in which such knowledge exists outside the mind of any single knower, and from which a knower can recover it.

The historical knowledge media are:
1) Oral tradition: Knowledge carried by speech and held, between utterances, in the memory of a community;
2) Written natural language: Knowledge carried by text written for human readers, from manuscripts to printed books and scientific papers.

However, recent times have seen the emergence of two other media:

\begin{description}
\item[LLM weights] Knowledge carried by the parameters of a trained language model and recovered by prompting it \cite{petroni2019language}.
\item[Code] Knowledge carried by text in a programming language or in a parsable data format such as JSON or TOML, and recovered both by reading the text and by executing it (\cref{sec:code-data}).
\end{description}

\subsection{Properties}

\Cref{tab:properties} lists thirteen properties for knowledge media.
Each property is defined independently of any medium and comes with a test question.
We group them in five families.

\begin{table*}
\caption{Epistemic properties of knowledge media. A medium has a property when the evidence answers the test question positively.}
\label{tab:properties}
\small
\begin{tabularx}{\textwidth}{@{}l l X@{}}
\toprule
Family & Property & Test question \\
\midrule
Keeping & Persistence & Does the knowledge exist when no one holds it in mind? \\
 & Transmissibility & Can it be copied from one holder to another without loss? \\
\midrule
Meaning & Precision & Do two competent readers recover the same content? \\
 & Expressiveness & How much of what we know can the medium state? \\
\midrule
Handling & Inspectability & Can a reader locate a given piece of knowledge and read it? \\
 & Editability & Can one piece be corrected while the rest stays unchanged? \\
 & Provenance & Is it recorded where each piece came from and when it changed? \\
 & Composability & Can a piece build on another piece by reference, without restating it? \\
\midrule
Executing & Executability & Can the medium itself, without a human interpreter, do what the knowledge describes? \\
 & Verifiability & Can a machine check whether a piece of knowledge is valid? \\
 & Reproducibility & Can the check be repeated later, by anyone, with the same procedure? \\
\midrule
Growing & Generativity & Is the medium also an instrument for obtaining new knowledge? \\
 & Scalability & At what rate can knowledge in the medium be produced and checked? \\
\bottomrule
\end{tabularx}
\end{table*}

\paragraph{Keeping}
Persistence and transmissibility are the properties that writing brought to oral tradition.
Writing turns speech into a durable object \cite[p.~11]{ong1982orality}, and print made identical copies of texts available in large numbers \cite[vol.~1, pp.~71--107]{eisenstein1979printing}.
They are necessary for a body of knowledge to outlive its authors.

\paragraph{Meaning}
Precision and expressiveness pull in opposite directions.
Natural language can state almost anything that we know, at the cost of admitting several readings of the same sentence.
Many formal languages admit one reading at the cost of restricting what can be stated.
No medium is complete: as Davis et al.\ put it, ``the only completely accurate representation of an object is the object itself'' \cite[p.~18]{davis1993knowledge}.
Expressiveness is therefore a matter of degree, and a medium has it when it can state most of what we know.

\paragraph{Handling}
Inspectability, editability, provenance and composability matter once a body of knowledge is large and maintained over time.
They decide whether one can find a piece of knowledge, fix it when it is wrong, know where it came from, and reuse it in new knowledge.

\paragraph{Executing}
Executability, verifiability and reproducibility concern the relation between a medium and the world it describes.
An executable medium performs what it describes.
It is knowing ``how'', in Ryle's sense \cite{ryle1949concept}, written down in a form that runs.
A verifiable medium supports a mechanical check of validity, without relying on the judgement of a reader.
Reproducibility asks whether that check can be repeated later, by anyone.
We agree with Claerbout who states that an article ``is not the scholarship itself, it is merely advertising of the scholarship. The actual scholarship is the complete software development environment and the complete set of instructions which generated the figures'' \cite{buckheit1995wavelab}.

\paragraph{Growing}
Generativity asks whether the medium is only a store of knowledge or also an instrument for obtaining it.
In experimental science, new knowledge comes from intervening in the world with instruments \cite{hacking1983representing}, and these instruments are made of something other than the knowledge they produce: glass, magnets, detectors.
A generative medium is one in which the instrument and the result are made of the same material.
Wang et al.\ show that language model agents perform better when they write their actions as ``executable Python code'', which they run and then revise ``upon new observations'' \cite{wang2024codeact}.
Such an action is an instrument written in the same medium as the knowledge it yields.
Scalability asks at what rate knowledge can be produced in the medium and checked, which bounds how large a body of knowledge can grow.

\subsection{Central claim}

\begin{quote}
\textbf{Knowledge as code.} Code is the first knowledge medium that has all thirteen properties.
\end{quote}

\Cref{sec:comparison} will fill the four media aforementioned.

\section{Knowledge as Code}
\label{sec:kac}

This section defines knowledge as code, and describes how it is produced, checked and revised.

\subsection{Definition}
\label{sec:definition}

\begin{definition}[Knowledge as code]
Knowledge as code is knowledge written in a programming language or in a machine-parsable data format, whose validity can be checked by executing it.
\end{definition}

A climate model encodes the laws of fluid dynamics and thermodynamics, and states how the temperature of the planet responds to greenhouse gases.
A radiotherapy planning system encodes how radiation deposits energy in human tissue, and states the dose that each organ receives.
A finite-element solver used in structural engineering encodes how materials deform under load, and states whether a bridge design holds.
Each is a claim about the world that can be executed, and the result of the execution can be compared with what the world does.

The knowledge in such code has two origins.
Part of it was first written in natural language, in textbooks, scientific papers and standards.
The code makes it executable, as the climate model executes equations that physics textbooks state in prose and formulas.

The other part is obtained by executing code against the world.
A climate model contains parameters that no textbook gives, and modelers obtain them by running the model.
``Tuning consists of adjusting the values of these parameters to bring the solution as a whole into line with aspects of the observed climate'' \cite{hourdin2017tuning}.
The tuned values were not written anywhere before the runs.
\Cref{sec:strata} elaborates on the two origins.

\textbf{The medium checks itself}
The clause ``whose validity can be checked by executing it'' is what separates code from the other knowledge media (see \cref{sec:properties}).
In written natural language, a claim and its check are different things.
The claim is a sentence; the check is an experiment, a proof or a reading, made by someone or something outside the text. 

In code, the knowledge and its check are made of the same material.
A finite-element solver is checked by running it on a structure whose deformation was measured, and a test is a program that executes other code and compares the result with an expectation written in the test itself.
Checking knowledge as code therefore needs no human reader and nothing outside the code, only a machine that executes it, which is why code has the properties of verifiability and reproducibility of \cref{tab:properties}.

\subsection{How Knowledge Enters Code}
\label{sec:strata}

Knowledge and learning are two views of one process.
Learning is how knowledge comes to be, and knowledge is what learning leaves behind.
A knowledge medium is therefore characterized not only by how it holds knowledge, the subject of \cref{sec:properties}, but also by how knowledge enters it.
In written natural language, knowledge enters the text after it was obtained elsewhere, in an experiment, a field study or a proof, and the text records the result without taking part in obtaining it.

In code, knowledge enters in two ways, via encoding and learning by execution.
Encoding moves existing knowledge into code; learning by execution obtains new knowledge.

\paragraph{Encoding Knowledge}
\begin{definition}[Encoded knowledge]
Encoded knowledge is knowledge that existed in written form before the code, for example in natural language, and that the code makes executable.
\end{definition}
Encoding moves into code knowledge that was first written elsewhere: a textbook, a scientific paper, a standard, a specification.
The knowledge changes medium, and in the new medium it can be executed and checked, hence gaining the epistemic properties that its written source lacks: precision, executability, verifiability and reproducibility (\cref{tab:properties}).

A numerical library turns the algorithms of numerical analysis textbooks into routines that any program can call and test.
At a smaller scale, the \texttt{sed} edits from our dataset encode the basic regular expressions of the POSIX standard (trajectory \texttt{390d0d04}).

\paragraph{Learning by execution}

\begin{definition}[Knowledge learned by execution]
Knowledge learned by execution is knowledge that did not exist in written form before, obtained by executing code against the world and observing the result, and written into code.
\end{definition}

It can enter any part of the code: a constant, a data structure, a function. Two examples follow, one from automatic performance tuning and one from an agent trajectory.
\begin{itemize}
\item ATLAS times many candidate implementations of linear algebra operations on the machine at hand, at install time, and builds the library from the fastest \cite{whaley2001atlas}.
The timing experiments are the probes, and the generated library is the knowledge they produce; on another machine, the experiments give another library.
FFTW does the same for Fourier transforms: its planner ``works by measuring the actual run time of many different plans and by selecting the fastest one'' \cite{frigo2005fftw3}.
\item An agent automating a messaging application (Telegram) runs a short Python program that creates a channel through the API and prints the response (trajectory \texttt{b8ea8c89}).
It reads the shape of the response, a list of updates of which one carries the identifier of the new channel.
It then writes a parser that walks that list and extracts the identifier, so the shape of the response is learned from the response.
\end{itemize}

\begin{lstlisting}[float=tp, caption={Knowledge as code in radiotherapy (synthetic and simplified: one attenuation coefficient stands for the depth-dose tables of real planning systems). The probe (lines 1--2) is run once; lines 4--16 are persisted. A monitor unit is the unit in which the machine counts the radiation it delivers; Gy is the unit of dose.}, label=lst:dose]
# Probe, run once against the treatment machine
print(detector.read(depth_cm=10, monitor_units=100))  # printed 0.667 (Gy)

import math
# Encoded: attenuation of 2 MeV photons in water (NIST, Hubbell and Seltzer)
MU_WATER = 0.0494  # per cm
# Learned by execution: probe of 2026-10-20 (line 2)
OUTPUT_GY_PER_MU = 0.667 / 100 / math.exp(-MU_WATER * 10)

def dose(depth_cm, monitor_units):
    """Dose in Gy on the axis of the beam, at a depth in water."""
    return OUTPUT_GY_PER_MU * monitor_units * math.exp(-MU_WATER * depth_cm)

def test_dose_matches_detector():
    measured = detector.read(depth_cm=10, monitor_units=100)
    assert abs(dose(10, 100) - measured) / measured < 0.02
\end{lstlisting}

\paragraph{Two roles of code}
Code plays two roles in learning by execution.

\begin{definition}[Probe]
A probe is code written to ask the world a question, and executed to obtain the answer as output, which is an observation.
\end{definition}

In the ATLAS and Telegram examples, the probes are the timing experiments of ATLAS and the program that creates a Telegram channel.

\begin{definition}[Persisted code]
Persisted code is code kept in a project to be executed again: its programs, libraries, configuration and tests, whatever the origin of the knowledge they hold.
\end{definition}

Persisted code holds encoded knowledge as well as knowledge learned by execution.
When it follows a probe, it captures what the probe revealed, as the generated ATLAS library and the Telegram parser do.
It then turns an observation, which held at one moment, into a claim that is executed and checked at every later use (\cref{sec:stable}).

The probe is the experimental means, and the persisted code is the experimental outcome.
Both are written in the same medium, which is why code is generative in the sense of \cref{sec:properties}.

\Cref{lst:dose} makes the radiotherapy example of \cref{sec:definition} concrete, in a synthetic and simplified form.
It computes the dose of radiation that a treatment machine delivers at a given depth in water, which stands for human tissue.
The attenuation coefficient of line 6 is encoded knowledge, copied from a physics reference table.
The output of this particular machine is knowledge learned by execution.
The probe of line 2 reads a radiation detector placed in the beam, and line 8 turns its observation into a constant.
The test of lines 14--16 runs the same measurement again and checks that the computed dose still agrees with it.
The attenuation of radiation in water is a law of physics and does not change, whereas the output of a machine drifts over time.
When the output drifts, the test fails, so the drift is detected by execution.

\paragraph{One single material}
Learning by execution is the meta-circular case where the instrument of learning and the record of what was learned are made of the same material: code.

What was learned can then be read, run and checked by anyone, so it has the properties of the medium that records it: inspectability, editability, provenance, verifiability and reproducibility.
Learning by execution is the case that the rest of this paper studies, and \cref{sec:results} measures how often agent-written code is learned this way.

\subsection{The Executable Knowledge Lifecycle}
\label{sec:lifecycle}

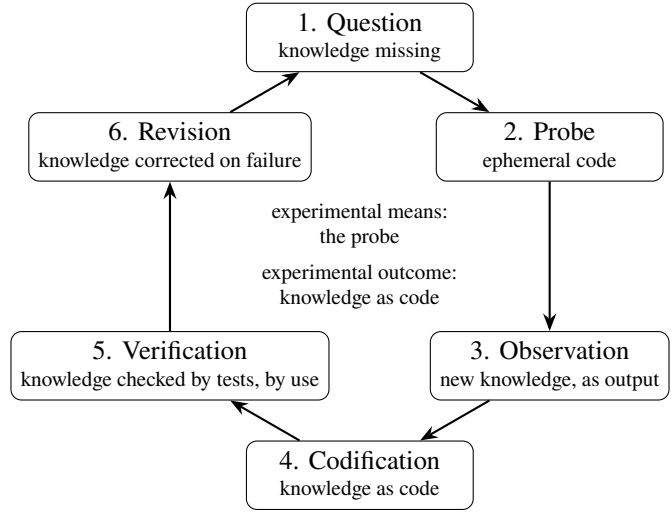
\begin{figure}
\centering
\begin{tikzpicture}[
  stage/.style={draw, rounded corners, minimum width=3.0cm, minimum height=0.9cm, align=center, font=\small},
  arr/.style={-{Stealth}, thick}]
\node[stage] (q) at (90:2.9) {1. Question\\[-2pt]{\scriptsize knowledge missing}};
\node[stage] (p) at (30:2.9) {2. Probe\\[-2pt]{\scriptsize ephemeral code}};
\node[stage] (o) at (-30:2.9) {3. Observation\\[-2pt]{\scriptsize new knowledge, as output}};
\node[stage] (c) at (-90:2.9) {4. Codification\\[-2pt]{\scriptsize knowledge as code}};
\node[stage] (v) at (-150:2.9) {5. Verification\\[-2pt]{\scriptsize knowledge checked by tests, by use}};
\node[stage] (r) at (150:2.9) {6. Revision\\[-2pt]{\scriptsize knowledge corrected on failure}};
\draw[arr] (q) -- (p);
\draw[arr] (p) -- (o);
\draw[arr] (o) -- (c);
\draw[arr] (c) -- (v);
\draw[arr] (v) -- (r);
\draw[arr] (r) -- (q);
\node[font=\scriptsize, align=center] at (0,0) {experimental means:\\the probe\\[4pt]experimental outcome:\\knowledge as code};
\end{tikzpicture}
\caption{The executable knowledge lifecycle. Probes and persisted code are both code, so the whole cycle, from missing knowledge to corrected knowledge, runs in one medium.}
\label{fig:lifecycle}
\end{figure}

Knowledge learned by execution is produced and maintained in a cycle of six stages, shown in \cref{fig:lifecycle}.
Each stage relies on one or more of the epistemic properties of \cref{tab:properties}.
\begin{enumerate}
\item \emph{Question.} A developer, human or agent, needs to know something about the world that is not written anywhere.
\item \emph{Probe.} The developer writes code that asks the world, often an ephemeral program that is run once and never saved \selfcite{monperrus2026ephemeral}.
The probe is possible because the medium is generative, so the instrument is written in the medium itself.
\item \emph{Observation.} The world answers with output, which the developer reads.
\item \emph{Codification.} The developer or coding agent writes what was learned into persisted code.
The scale ranges from a single constant to a function, a module, a protocol implementation or a whole system.
Once persisted, the knowledge has persistence and transmissibility, and it is composable, since other code imports and calls it.
\item \emph{Verification.} The persisted code is executed again, in two ways.
Verification by tests runs code written to check the knowledge, which is verifiability.
Verification by use runs the knowledge itself on the task it was written for, which is executability.
Both can be repeated by anyone at any later time, which is reproducibility.
\item \emph{Revision.} A failed verification shows that the code, or the world, is not as believed.
The code is changed, which relies on editability, and version control records what changed and when, which is provenance.
The cycle starts again.
\end{enumerate}

\subsection{Stability of Knowledge-as-code}
\label{sec:stable}

The output of a probe and the assertion of a test do not have the same status.
The output of a probe is not code.
It is an observation, text printed once, which records the state of the world at one moment, such as the files in a directory or the response of an API on a given day.
It is contingent, since tomorrow it can differ.
Codification turns such observations into code that is meant to hold over time.

\begin{definition}[Stable code knowledge]
Stable code knowledge is knowledge stated by persisted code that is expected to hold whenever the code is executed, so that its check passes today and at any later time.
\end{definition}

How stable a piece of code knowledge is depends on what it is about.
The equations that a climate model or a finite-element solver encodes are laws of physics, and they do not change, so this code knowledge is among the most stable there is.
The tuned parameters of the climate model hold as long as the observations they were fitted to stay the reference.
The library that ATLAS generates only holds for the machine it was tuned on.
Encoded knowledge inherits the stability of its source, and knowledge learned by execution inherits the stability of the part of the world that was probed.

Tests are the instrument that measures stability.
An assertion states what the code under test must do.
Running the test suite again, a day or a year later, checks whether the code still does it, without a human reader and without any information beyond the code and the world it runs against.
A test suite that passes again is evidence that the knowledge it checks is stable.
A test that fails after the world changed, with no change to the code, marks knowledge that was contingent although it was written as stable, and starts a revision.
No other medium carries, next to each claim, the procedure that checks whether the claim still holds.
\Cref{sec:results} runs this check on the test suites of agent-written repositories.

Refreshing knowledge as code follows a cycle.
A test fails, a new probe asks the world what changed, the answer is written as new knowledge as code, and a new test verifies it.
Each turn of this cycle replaces knowledge that no longer holds with knowledge checked against the world as it is now.
So, learning by execution with probe is the primary way to refresh knowledge: a probe of the installed library replaces what the weights remember with what the probe shows about the world.

\section{Validation}
\label{sec:validation}

\subsection{Experimental Method}
\label{sec:method}

We measure the concepts of \cref{sec:kac} based on the work of coding agents.
Each research question observes one part of the executable knowledge lifecycle of \cref{fig:lifecycle}.
\begin{description}
\item[RQ1 Volume] How much of what agents write is knowledge as code, and how much of it is probes, as opposed to persisted code? (stages 2 and 4)
\item[RQ2 Learning by execution] How much persisted code is knowledge learned by execution, and how much is encoded knowledge? (stages 3 and 4)
\item[RQ3 Verification] How much persisted code is test code, and do tests check knowledge learned by execution? (stage 5)
\item[RQ4 Stability] Does the knowledge still hold months later? (stages 5 and 6)
\end{description}

\subsubsection{Corpus}
\label{sec:corpus}

A coding agent is a language model that writes code and executes it, in a loop, through tools.
At each step, it emits an action, such as a shell command, a file edit or a program to run, and reads the output of that action before choosing the next one \cite{yao2023react,wang2024codeact}.
Its session is recorded in a \emph{trajectory}, the log of every action and every observation.
A human developer who tries a command before writing code leaves no trace of the command, while an agent does.
This is why we study knowledge as code through coding agents, since their trajectories record the probe, the observation and the persisted code together.

The corpus consists of the trajectories of the coding agents used by the author, a single developer, from February to October 2026.
It contains 29,260 trajectories and 462,448 tool calls.
The harnesses are \harness\footnote{\harnessurl} (16,068 trajectories), Claude Code\footnote{\url{https://github.com/anthropics/claude-code}} (10,792), Copilot CLI\footnote{\url{https://github.com/github/copilot-cli}} (1,200), Codex\footnote{\url{https://github.com/openai/codex}} (673), OpenCode\footnote{\url{https://github.com/anomalyco/opencode}} (503) and Hermes\footnote{\url{https://github.com/NousResearch/hermes-agent}} (24).
The trajectories come from the author's laptop and from two servers.

A trajectory records the first four stages of the lifecycle.
The question is in the prompt and in the messages of the agent, every probe is a tool call whose output is recorded, and every write of persisted code is a tool call whose arguments contain the code.
\subsubsection{Operational Definitions}
\label{sec:operational}

The definitions of \cref{sec:kac} become measurable as follows.
\begin{itemize}
\item \emph{Knowledge as code} is a line written in a file whose extension denotes a programming language or a parsable data format (JSON, YAML, TOML, configuration files).
Markdown, LaTeX and plain text are prose.
\item A \emph{probe} is a shell command or an inline program whose output the agent reads.
A probe is a \emph{run} when it executes a program, such as a script, a test suite or a compiler, and prints no existing file.
It is a \emph{read} when one of its commands prints existing files, as \texttt{cat}, \texttt{grep} and \texttt{git show} do, since its output then mixes existing text with observation.
\item An \emph{ephemeral probe} is a program piped to an interpreter and never saved, as in \texttt{python3 -\,<<EOF} \selfcite{monperrus2026ephemeral}.
\item \emph{Persisted code} is a write or an edit of a code or data file: a file written whole, the new text of an edit, the added lines of a patch, or a shell heredoc redirected to a file.
\item \emph{Knowledge learned by execution} is persisted code that contains a distinctive token first revealed by a run (\cref{sec:rq2}).
\item \emph{Encoded knowledge} is persisted code that contains no such token.
\end{itemize}

\subsubsection{RQ1: Volume}
\label{sec:rq1}

We count every line of code, structured data and prose that agents wrote through a tool call, with the \texttt{count-loc} command of \counter, the counter of \selftextcite{monperrus2026ephemeral}.
A write counts all its lines, an edit counts the lines of its new text, and a rewrite counts again each time it is made.
Failed tool calls are dropped, and calls are deduplicated by identifier.
The counter assigns each line to a medium (code, structured data, prose) and to a location: the project directory, the scratch directory, elsewhere on disk, or an ephemeral probe.
We count over all harnesses and hosts, 80,207 authored chunks in total.

\subsubsection{RQ2: Learning by Execution}
\label{sec:rq2}

We link each persisted code edit to the probes that preceded it in the same session.
The link is computed automatically, on the Claude Code sessions, which hold 76.5\% of the code lines of RQ1: 7,929 sessions in 402 projects, with 24,957 persisted edits of code or data files.

\paragraph{Distinctive tokens}
A distinctive token is a number of three or more digits, a decimal number, or an identifier of eight or more characters that contains a digit, a separator or mixed case, such as \texttt{channel\_id}, \texttt{/bin/pg\_dump} or a hexadecimal address.
Plain words never count.
Comments, docstrings and markup text are removed from the edit with a lexer before matching, so only the code of the edit is compared.

\paragraph{Link}
For every token of the session, we record where it appeared first: in the output of a run, in the output of a read, or anywhere else (the prompt, a message, a file read through the editor, code the agent wrote before).
A persisted edit is learned by execution when its code contains a token that first appeared in the output of a run.
It is encoded from existing code on disk when its tokens first appeared only in the output of reads.
It is encoded from the agent's prior knowledge otherwise.
Prior knowledge covers both what the weights hold and what the agent read through tools other than the shell; this study does not tell the two apart.

\paragraph{Prior-knowledge filter}
A token that the agent writes in ten or more distinct projects, such as \texttt{json.dumps} or \texttt{argparse.ArgumentParser}, belongs to its prior knowledge, not to what a session observed, and never links.
We report the results for thresholds of 2 to 50 projects.

\paragraph{Null model}
To measure links that arise by chance, we pair each session with the next session, in identifier order, from another project, and count the edits whose code contains a non-common token that appears in the run outputs of the paired session.
This condition ignores order and first appearance, so it overestimates chance.

\subsubsection{RQ3: Verification}
\label{sec:rq3}

Test code is persisted code in a file that follows the path conventions of a test framework: \texttt{test\_*.py} and \texttt{conftest.py} for pytest, \texttt{*\_test.go} for Go, \texttt{*.test.ts} and \texttt{*.spec.js} for JavaScript runners, \texttt{*Test.java} for JUnit, \texttt{*.t.sol} for Foundry, and any file under a \texttt{test} or \texttt{tests} directory.
We measure the share of persisted code lines that are test code, ephemeral probes excluded.
We then compare, with the link of RQ2, the share of test edits and of other edits that are learned by execution.
A test that asserts an observed value checks knowledge learned by execution.

\subsubsection{RQ4: Stability}
\label{sec:rq4}

Stability is measured on observations and on persisted code.

\paragraph{Observations}
We replay past probes today and compare their output with the recorded one.
The candidates are the runs of the Claude Code sessions of the laptop, the host where the replay takes place.
We exclude probes that need the network (16,074), probes that use a session's scratch directory, which no longer exists (2,747), probes that act on the system, such as \texttt{kill} or \texttt{systemctl} (932), probes run in the background (815), and probes whose recorded output was truncated (229).
Of the 18,067 remaining probes, we draw a random sample of 400 (seed 2026).

Each probe runs twice in a sandbox built with bubblewrap: the whole file system read-only, empty and disposable \texttt{/tmp}, \texttt{/run} and cache directories, no network, and separate process namespaces.
A probe cannot change the machine or reach the outside world.
We compare outputs with two oracles, in this order.
The first is output equality after normalizing dates, times, durations, identifiers, process numbers, temporary paths and the notes that the harness adds to outputs.
The second is an LLM judge (Sonnet 5.5), asked whether the replay gives the same answer as the recorded output to the question the command asks, and whether a difference comes from a changed world or from a replay environment that cannot answer, for instance because a credential is missing.
The author checked a random sample of 20 verdicts of the judge and agreed with all of them.
A replay that fails because of the sandbox itself, such as a write to the read-only file system, is set apart and says nothing about the world.

\paragraph{Persisted code}
Tests are the instrument that measures stability, so we run the latest version of test suites today.
We restrict this to repositories written entirely by agents, so that the result concerns agent-written knowledge.
A repository qualifies when every non-blank code line of its working tree appears in an agent edit of that file, in a trajectory of any harness.
Edits include whole-file writes, string replacements, patches, heredocs and in-place \texttt{sed} edits, whose substitutions we apply to the lines the agent had written before.
We scan the 942 git repositories of the author's workspace, find the test command of each qualifying repository from its build files (pytest, \texttt{go test}, \texttt{cargo test}, \texttt{npm test}, \texttt{make test}), and run the suite once in the same sandbox, with build and cache directories made writable and disposable, and a limit of ten minutes.
A suite that passes again measures that the knowledge it checks is stable.

\subsubsection{Instrument Validation}
\label{sec:instrument-validation}

Every instrument is tested on a known-positive, a known-negative and a boundary case, 62 unit tests in total.
RQ1 reuses the counter of \selftextcite{monperrus2026ephemeral}, which publishes 1,362,883 lines of code for an earlier snapshot of the same corpus; our total of 1,364,208 lines is within 0.1\% of it.
RQ2 is checked by its null model and by the sensitivity of its result to the prior-knowledge threshold.
The replay sandbox of RQ4 is checked to block writes and network access, and its judge by the author's review.
The instruments and their raw results are part of the replication package.

\subsection{Experimental Results}
\label{sec:results}

Each research question ends with what its results say about the epistemic properties of \cref{tab:properties}.

\subsubsection{RQ1: Volume}

Coding agents wrote 1,741,339 lines through tool calls.
Of these, 1,364,208 are code (78.3\%), 38,436 are structured data (2.2\%), 331,262 are prose (19.0\%), and 7,433 have no recognizable format (0.4\%).
Knowledge as code, code and structured data together, therefore makes up 80.6\% of what agents write, 4.2 lines for every line of prose.
\Cref{tab:rq1} shows that this holds for every harness, between 58.2\% and 89.3\%.

\begin{table}
\caption{RQ1: lines written by coding agents, by harness.}
\label{tab:rq1}
\small
\begin{tabular}{@{}lrrrr@{}}
\toprule
Harness & Code & Data & Prose & Knowledge as code \\
\midrule
\textbf{All} & \textbf{1,364,208} & \textbf{38,436} & \textbf{331,262} & \textbf{80.6\%} \\
\midrule
Claude Code & 1,043,909 & 34,026 & 289,961 & 78.5\% \\
\harness & 265,213 & 3,209 & 31,566 & 89.3\% \\
Copilot CLI & 48,390 & 605 & 5,722 & 88.8\% \\
OpenCode & 3,956 & 527 & 3,140 & 58.2\% \\
Codex & 2,740 & 69 & 873 & 76.3\% \\
\bottomrule
\end{tabular}
\end{table}

Probes are a large part of this code.
Of the 1,364,208 lines of code, 351,518 (25.8\%) are ephemeral probes, programs that were executed once and never saved.
One line of code in four is therefore written to ask the world a question, not to keep an answer.

\paragraph{Epistemic properties}
The volume shows scalability in practice, since the agents of a single developer produced 1.4 million lines of knowledge as code in eight months.
The probes show generativity in practice, since a quarter of the code is not knowledge kept but an instrument for obtaining knowledge, written in the same medium.

\begin{rqanswer}{RQ1}
Four lines out of five that coding agents write are knowledge as code.
One line of code in four is a probe, an instrument written in the medium itself.
Code is a scalable and generative knowledge medium in practice, not only in principle.
\end{rqanswer}

\subsubsection{RQ2: Learning by Execution}

Of the 24,957 persisted edits of code or data files in Claude Code sessions, 2,900 (11.6\%) contain a token first revealed by a run, and thus are knowledge learned by execution.
They account for 211,599 of the 759,103 edited lines (27.9\%), since edits learned by execution are longer than average.
A further 7,831 edits (31.4\%) contain tokens first revealed by reads only, and are knowledge encoded from existing code on disk.
The remaining 14,226 edits (57.0\%) are encoded from the agent's prior knowledge in the model.

Learning by execution is not an artifact of chance.
The null model, which matches each session against the run outputs of an unrelated session, links 0.12\% of edits, two orders of magnitude below the 11.6\% of the real sessions.
The share stays within a factor of two of 11.6\% across thresholds of the prior-knowledge filter.
It is 7.2\% when a token written in two projects is already prior knowledge, 10.5\% at five projects, 12.5\% at twenty and 13.5\% at fifty.
Without the filter, 14.7\% of edits are linked to a run.
Since a link requires a shared token, these numbers are a lower bound: code written in response to an observation without copying any token from it, such as an error handler added after a failure, is not counted.

\paragraph{Epistemic properties}
Knowledge learned by execution illustrates generativity in two ways, since the probe produces knowledge and the medium of the probe is the medium of the result.
The link itself shows provenance.
Because both the probe and the persisted code are code, recorded in the same trajectory, a line of code can be traced back to the very observation that produced it.

\begin{rqanswer}{RQ2}
At least 27.9\% of the persisted code lines that agents write are knowledge learned by execution.
This is a hundred times more than chance.
Each of these lines can be traced to the observation that produced it, which demonstrates the provenance epistemic property.
\end{rqanswer}

\subsubsection{RQ3: Verification}

Of the 1,012,690 lines of persisted code, 210,625 (20.8\%) are test code.
The share varies by language: 16.9\% in Python, 25.8\% in Go, 31.2\% in JavaScript, 37.9\% in Java, and 66.7\% in Solidity, where the agents of the corpus study deployed contracts through tests that run against a local copy of the blockchain.
Without Solidity, the share is 16.9\%.

Test code holds knowledge learned by execution twice as often as other code.
Of the 24,957 edits of RQ2, 23,077 edit code files, and the split below covers them.
The other 1,880 edit data files, which are neither tests nor programs.
Of the 2,999 edits of test files, 572 (19.1\%) are learned by execution, 982 (32.7\%) are encoded from existing code on disk, and 1,445 (48.2\%) are encoded from the agent's prior knowledge.
Of the 20,078 edits of other code files, 1,902 (9.5\%) are learned by execution, 6,329 (31.5\%) are encoded from code on disk, and 11,847 (59.0\%) from prior knowledge.
The share encoded from code on disk is the same in tests and in other code.
The difference lies between the two other outcomes, since tests trade prior knowledge for observed values, which is the codification stage of the lifecycle feeding the verification stage.

\paragraph{Epistemic properties}
Tests are verifiability made concrete, since one fifth of the persisted code exists only to check the rest.
Verifiability is applied first to knowledge learned by execution.
Knowledge learned by execution comes from contingent observations, so it is the knowledge most exposed to a changing world, and the knowledge whose check matters most.
A test makes the verifiability check reproducible, since anyone can run it again later.

\begin{rqanswer}{RQ3}
One line of persisted code in five is a test.
An edit of a test is twice as likely as an edit of other code to contain a value that a probe observed (19.1\% against 9.5\%).
Verifiability is applied first to the knowledge that the world taught.
\end{rqanswer}

\subsubsection{RQ4: Stability}

\paragraph{Observations}
\Cref{tab:rq4} gives the outcome of replaying the 400 sampled probes.

\begin{table}
\caption{RQ4: replay of 400 probes sampled from past trajectories.}
\label{tab:rq4}
\small
\begin{tabular}{@{}lr@{}}
\toprule
Outcome & Probes \\
\midrule
Same output after normalization (first oracle) & 61 \\
Same answer, different output (judge) & 17 \\
World changed (judge) & 31 \\
Replay environment cannot answer (judge) & 110 \\
Blocked by the sandbox & 130 \\
Timeout & 29 \\
Working directory gone & 22 \\
\midrule
Total & 400 \\
\bottomrule
\end{tabular}
\end{table}

Under the oracle of normalized output equality, the output of 61 of the 219 probes that ran to completion is the same as months before (27.9\%).
Raw observations rarely repeat exactly, and this is expected, because
the output of a probe captures a transient state of the world, such as the current size of a file, the number of entries in a log or the last commit of a repository, and that state has moved on since the recording.
The output of a probe is contingent, as \cref{sec:stable} states, and it is not yet knowledge as code.

The judge finds that 17 more replays give the same answer in a different form, that 31 reveal a changed world, and that 110 differ because the replay environment cannot answer.
Among the 109 probes whose replay is informative about the world, what the probe revealed still holds for 78 (71.6\%).
Most of what probes reveal is thus about a stable part of the world, and code that captures it remains valid.

For the other 31 probes (28.4\%), the world changed, in three ways: an installed tool or library was upgraded, such as a command-line tool that gained subcommands; a data file grew or changed, such as a dataset that gained entries; the probed code itself changed, such as a class that lost a method.

The share of probes whose knowledge still holds does not decrease with age within the eight months of the corpus: it is 66.7\% for probes less than 30 days old (8 of 12), 77.5\% between 30 and 90 days (31 of 40), and 68.4\% beyond 90 days (39 of 57).
What changes in the world changes for specific reasons, such as an upgrade, not by gradual decay.
Knowledge learned by execution is therefore not short-lived.

\paragraph{Reproducibility over time}
Of the 942 repositories of the author's workspace, 41 are written entirely by agents, 53 at 99\% or more, and 74 at 95\% or more.
Of the 41, 8 have a test command that can be found, and none of their test suites fails.

Five suites run their tests and pass, with 3 to 15 tests each, up to seven weeks after their last commit.
A sixth suite passes with all its tests skipped, since they modify a live mail account and run only on request.
The knowledge that these tests check has stayed valid, while the raw observations of the same period changed in 28.4\% of the informative cases.
This is the contrast that \cref{sec:stable} predicts: codification turns contingent observations into stable code knowledge, and the tests confirm it.

Two suites cannot run.
One writes into the repository, which the sandbox forbids, and the build tool of the other cannot start offline.
Neither failure says anything about the knowledge, only about the replay harness.
Finally, 33 of the 41 agent-written repositories have no test suite at all, so their knowledge cannot be re-checked except by use.

\paragraph{Epistemic properties}
RQ4 measures reproducibility directly: a check made months ago is repeated today, by a machine, with the same procedure.
For knowledge as code, the check holds in most of the cases that can be repeated.
Its limit is the changing environment.

\begin{rqanswer}{RQ4}
What a probe revealed months ago still holds for 71.6\% of the replays that can be checked.
This share does not decrease over the eight months of the corpus.
Every agent-written test suite that runs still passes, so the reproducibility of knowledge as code holds over time.
\end{rqanswer}

\subsection{Threats to Validity}
\label{sec:threats}

\paragraph{Construct validity}
Each measure approximates a concept of \cref{sec:kac}.
\begin{description}
\item[RQ1] Knowledge as code is recognized by file extension.
Code embedded in Markdown or in notebooks is missed, and a data file with an unusual extension is counted as unknown.
\item[RQ2] The link requires a token shared between an observation and a code edit.
It misses knowledge learned by execution that leaves no token, such as an error handler added after a failure, so its 11.6\% is a lower bound.
A command that both runs a program and prints a file is counted as a read, which moves knowledge learned by execution towards encoded knowledge.
\item[RQ3] Test code is recognized by path conventions.
Tests stored elsewhere are missed, and helper files of test directories are counted as tests.
\item[RQ4] The normalization of outputs is incomplete, and the judge can err.
The provenance check of agent-written repositories matches lines within a file, so a trivial line such as a closing brace can be attributed to an agent by coincidence, and renamed files are not followed.
\end{description}

\paragraph{Internal validity}
The agents worked under the standing instructions of the author, which ask for tests and for executable checks of results.
The share of test code in RQ3 reflects these instructions, and is an upper estimate for developers who give none.
The sandbox of RQ4 excludes probes that need the network, and probes with side effects.
The sample therefore over-represents knowledge about the local machine, and what holds for it may not hold for knowledge about remote services, which change at the will of their providers.
The judge was validated by the author.

\paragraph{External validity}
The corpus comes from one developer, a researcher, over eight months, on research prototypes, with three quarters of the code in Python.
The proportions measured here are not established for industrial teams, other domains or other languages.
The trajectories contain private data and cannot be released, which limits replication to the instruments.
However, the instruments are open source and can be run on any trajectory, so the study can be repeated on other corpora.

\section{Comparing Knowledge Media}
\label{sec:comparison}

This section applies the thirteen properties of \cref{tab:properties} to the four media of \cref{sec:properties}: oral tradition, written natural language, LLM weights and code.
For each medium and property, the verdict is \emph{yes}, \emph{partly}, or \emph{no}, and each verdict rests on a citation or on the nature of the medium.
\Cref{tab:comparison} gives the grid, and the subsections below justify it, one family of properties at a time, ending each with the case of code.

\begin{table}
\caption{The thirteen epistemic properties of \cref{tab:properties}, for four knowledge media.}
\label{tab:comparison}
\small
\begin{tabular}{@{}llcccc@{}}
\toprule
Family & Property & Oral tradition & Written language & LLM weights & Code \\
\midrule
Keeping & Persistence & no & yes & yes & yes \\
 & Transmissibility & no & yes & yes & yes \\
\midrule
Meaning & Precision & partly & partly & no & yes \\
 & Expressiveness & yes & yes & yes & yes \\
\midrule
Handling & Inspectability & no & yes & no & yes \\
 & Editability & no & yes & partly & yes \\
 & Provenance & no & yes & no & yes \\
 & Composability & no & yes & partly & yes \\
\midrule
Executing & Executability & no & no & no & yes \\
 & Verifiability & no & no & partly & yes \\
 & Reproducibility & no & partly & partly & yes \\
\midrule
Growing & Generativity & no & no & partly & yes \\
 & Scalability & no & partly & partly & yes \\
\midrule
\multicolumn{2}{@{}l}{Number of \emph{yes}} & 1 & 7 & 3 & 13 \\
\bottomrule
\end{tabular}
\end{table}

\subsection{Persistence, Transmissibility}

Oral tradition keeps knowledge only in living memory.
``Since in a primary oral culture conceptualized knowledge that is not repeated aloud soon vanishes, oral societies must invest great energy in saying over and over again what has been learned arduously over the ages'' \cite[p.~41]{ong1982orality}.
Transmission changes what it transmits: genealogies are ``transmuted in the course of being transmitted'' \cite[p.~310]{goody1963consequences}, and oral societies keep their equilibrium ``by sloughing off memories which no longer have present relevance'' \cite[p.~46]{ong1982orality}.
Writing turns words into a ``residue'' that persists \cite[p.~11]{ong1982orality}, and print makes copies identical \cite[vol.~1, pp.~71--107]{eisenstein1979printing}.
LLM weights are files, which persist and are copied bit for bit.

\paragraph{Code}
Code keeps knowledge as reliably as print.
A file of code persists without anyone holding it in mind, and every copy is identical to the original, down to the last byte.
Code also has its library of Alexandria.
Software Heritage is ``the largest public archive of software source code and associated development history'', and ``as of July 2023, it has archived more than 16 billion unique source code files coming from more than 250 million collaborative development projects'' \cite[p.~33]{dicosmo2023heritage}.

\subsection{Precision, Expressiveness}

Natural language, spoken or written, can state almost anything that we know, and it admits several readings of the same sentence.
LLM weights are less precise still: a model gives different answers to paraphrases of the same question, and can hold contradictory beliefs \cite{alkhamissi2022review}.
On expressiveness, the weights do well, since they ``require no schema engineering'' and ``allow practitioners to query about an open class of relations'' \cite{petroni2019language}.

\paragraph{Code}
Code is the only medium of the four that is both precise and expressive.
A program has one reading, the one the machine executes, defined by the semantics of its language.
This precision does not cost expressiveness.
A general-purpose programming language is universal in Turing's sense, so any computation that can be described can be written as a program \cite{turing1937computable}.
What is not a computation is still carried by the code, in its identifiers, comments and documentation, next to what it computes.
A tax program computes the tax of each income bracket exactly, and the comment next to the rates can state the political choice behind them.

\subsection{Inspectability, Editability, Provenance, Composability}
Oral tradition has no artifact to inspect, edit or cite.
Written language has all four properties once it is digital and kept under version control.
A Wikipedia article or a paper written in LaTeX under git can be read by anyone, which is inspectability.
One sentence can be corrected while the rest stays unchanged, which is editability.
The history records who changed each sentence and when, which is provenance.
A text builds on another text by citation, without restating it, which is composability.

LLM weights have neither inspectability nor provenance.
``Knowledge in LMs is encoded in a diffused manner, making it generally difficult to interpret and hard to update'' \cite{alkhamissi2022review}, and the origin of a fact is lost in training.
Editability and composability hold only in part.
A single fact can be edited in the weights \cite{meng2022locating}, but the edit rarely reaches the facts that depend on it or composes with them \cite{cohen2024evaluating}.

\paragraph{Code}
Code has these four properties in a stronger form than writing.
A file can be read by any developer, which is inspectability, and changed line by line with a diff that touches nothing else, which is editability.
Version control records who changed what and when, which is provenance.
For knowledge learned by execution, provenance goes further, since the trajectory of an agent records the observation that a line of code came from.
Code shines with composition.
A reader who follows a citation must still understand the cited text before using it, whereas a program imports a library and executes it without anyone restating or reading it.

\subsection{Executability, Verifiability, Reproducibility}

Oral and written language need a human to carry out what they describe.
For written language, the check is an experiment made outside the text, which a reader can repeat with a procedure of their own.
LLM weights produce text, which acts on the world only once it is turned into code or read by a human.
A fact in the weights can be queried \cite{petroni2019language}, which is a mechanical check of what the model holds, but not of whether the world agrees.

\paragraph{Code}
Code is the only medium that acts, and the only one that can carry its own check.
Executing a program carries out what it describes, with no human in between.
A test is code that checks other code, which is verifiability.
The same test can be run again, a day or a year later, against the same code, which is reproducibility.

\subsection{Generativity, Scalability}

Oral and written language store knowledge obtained elsewhere, and grow at the pace of human speech and writing.
LLM weights generalize beyond their training data, but they learn by training, which adjusts parameters that no one can read one by one, and what they learn can only be queried.
They absorb knowledge without human supervision \cite{petroni2019language}, but accuracy on rare facts would require models ``scaled dramatically (e.g., to one quadrillion parameters)'' \cite{kandpal2023large}, and checking what they hold does not scale with them.

\paragraph{Code}
Code is the only medium that learns from the world in its own material, at machine speed.
A probe is an instrument written in the medium itself, and what the probe reveals is written back as persisted code.
A coding agent uses the two media at once.
Its weights hold what it learned by training, before the session started.
What it learns during a session, by encoding or by execution, goes into code, not into its weights.

\subsection{Epistemic properties of code}

The central claim of \cref{sec:properties} is that code is the first knowledge medium that has all thirteen properties.
\Cref{tab:comparison} supports it.

Code has all thirteen properties.
No other medium has them all.
Written language adds persistence, transmissibility and inspectability to oral tradition, and, once digital and under version control, editability, provenance and composability.
LLM weights keep persistence, transmissibility and expressiveness, lose inspectability and provenance, and keep editability only in part.

Code is the first medium ever to add executability, verifiability, reproducibility and generativity, and it does so without losing what writing brought. Code is the most sophisticated version of writing.

\subsection{Limits of knowledge as code}
\label{sec:limits}

Having the thirteen properties does not mean having each of them without limit.
Four properties have limits that matter for knowledge as code, namely expressiveness, inspectability, verifiability and reproducibility.

\paragraph{Limit to expressiveness}
Code can state any computation, but not all knowledge is a computation.
Goals, values and the reasons behind a choice cannot be executed.
This is why code supports comments.
Code carries uncomputable knowledge in natural language, in comments, identifiers and documentation, next to what it computes.
A comment that records why a safety margin was set to its value is an example.
Since comments and identifiers are part of the code, code states both what is computed and why.

\paragraph{Limit to inspectability}
Every line of code can be read, but an execution cannot always be followed.
Humphreys calls a process ``epistemically opaque relative to a cognitive agent X at time t just in case X does not know at t all of the epistemically relevant elements of the process'' \cite[p.~618]{humphreys2009novelty}.
A climate model that runs billions of steps is opaque in this sense, although its code is inspectable line by line.
Inspectability of code is thus inspectability of the text, and an execution is inspected only through what it records, such as its outputs, its logs and the results of its tests.

\paragraph{Limit to verifiability}
Code is verifiable, but only the parts that are executed and checked are verified.
The prose embedded in code persists and composes with the rest, but no execution checks it, so it has the verifiability of prose.
The executable part is verified only where tests exist.
Being code is what makes verification possible, and the discipline of writing tests is what puts it into practice.

\paragraph{Limit to reproducibility}
Reproducibility holds over time only as long as the knowledge stays true, and code knowledge can go stale, from two sources.
Knowledge that an agent encodes from its weights describes code as it was at the training cutoff, since language models predict text written after their training period worse than text written before it \cite{lazaridou2021mind}.
Knowledge that was learned by execution goes stale when the part of the world that was probed changes.
In both cases, a test detects it, since the knowledge that the test checks no longer holds and the test fails.
Knowledge staleness is thus not silent, as in a book or in weights, but detected by execution.
\Cref{sec:stable} describes how stale knowledge as code is refreshed.

\section{Discussion}
\label{sec:discussion}

\subsection{Knowledge of the Physical World}
\label{sec:physical}

Knowledge as code also holds for the physical world, to a significant extent.
Code reaches the physical world through sensors, which turn it into data, and through actuators, which act on it.
For example, in a study at Anthropic, Claude Code learned to control a quadruped robot by executing code against it, 19 times faster than the faster of the two human teams \cite{ilie2026projectfetch2}.
Knowledge learned by execution against the device was more reliable than knowledge found online, which misled one of the human teams \cite{anthropic2025projectfetch}.
The resulting control code is knowledge of the robot itself and about the world it is embedded in, which is executability in the physical world.

Verifiability holds in the physical world, through an instrument.
In the digital world, the probe and the object probed are made of the same material, since both are code and data on a computer or its network.
In the physical world, every probe is mediated by an instrument that turns the world into data \cite{hacking1983representing}, so a check of code knowledge is only as good as the instrument it reads.

Reproducibility holds in the physical world, at a cost.
A probe of a physical device has physical side effects and cannot be replayed in a sandbox as in RQ4.
Repeating a check costs time, material and sometimes safety, so code knowledge about the physical world is checked less often than code knowledge about the digital world, though it is checked the same way, by running code.
\subsection{Science as Code}
\label{sec:science-as-code}

Science's grand mission is to produce knowledge about the world.
\Cref{sec:comparison} shows that code is the knowledge medium with the most epistemic properties, and the only one that has them all.
If science produces knowledge, and code is its best medium, then science and code belong together, and a scientific result is best kept as code.

The classical analysis of knowledge asks for three things: a claim, its justification, and its truth \cite{steup2024epistemology}.
In science written in natural language, the three live in different places.
The claim is in the paper, the justification is in an experiment that was run elsewhere, and its falsifiable truth is checked by other laboratories, if ever.

In knowledge as code, the three live in one single medium.
The claim is the persisted code, its justification is the probe and its observation recorded in the trajectory, and its truth is checked by the test, which anyone can run again.
We call science as code the production of scientific knowledge in this form, where claim, justification and check are all code.
The trajectory keeps the justification of each claim as a laboratory notebook keeps an experiment.
Agents for science, such as Coscientist \cite{boiko2023coscientist} and The AI Scientist \cite{lu2024aiscientist,lu2026endtoend}, are agents entirely founded on science as code.

\section{Related Work}
\label{sec:related}

\subsection{Code as knowledge}
\textbf{Software as a knowledge medium.}
Armour states that software ``is a medium'', and ranks it after DNA, the brain, hardware and books: ``Software is the fifth knowledge storage medium'' \cite{armour2000case}.
For Armour, what sets software apart is that its knowledge is ``active'' and can ``be executed''.
He also observes that ``too often the activity of building a system is actually just the mechanism we use to acquire the knowledge'' \cite{armour2000case}.
This is the closest statement of our thesis, made in a column on the business of software.
We give it an epistemic foundation, with thirteen properties that compare code with other media, the two ways in which knowledge enters code, and measurements on real work.

\textbf{Expert systems.}
Knowledge engineering is ``the art of building complex computer programs that represent and reason with knowledge of the world'' \cite[p.~1016]{feigenbaum1977art}.
In an expert system, ``the interesting action arises from the knowledge base, not the inference engine'' \cite[p.~1018]{feigenbaum1977art}, so the knowledge is held by the program.
This knowledge comes from human experts, whose ``rules of expertise'' knowledge engineers ``seek to transfer to our programs'' \cite[p.~1016]{feigenbaum1977art}, which is encoding in the sense of this paper.
Knowledge as code adds the second origin, learning by execution, where the program obtains knowledge from the world instead of from an expert.

\textbf{World models as code.}
The closest system is WorldCoder, an agent that ``builds a Python program representing its knowledge of the world based on its interactions with the environment'' \cite{tang2024worldcoder}.
It learns by execution in the sense of this paper, and it keeps what it learns as code.
Its program is a model of the transitions and rewards of game-like environments, used for planning.
Our work takes the same mechanism as a general epistemic notion, compares code with other knowledge media, and measures it on real software work, where tests check the knowledge.

\textbf{Code as a knowledge representation.}
Gao proposes code annotated with its semantics, evidence and validation state as a knowledge representation for coding agents \cite{gao2026eck}.
It shares the view that code can carry knowledge with its own check, and it is an engineering proposal, not an epistemological one; it does not study how such knowledge is learned from the world.

\textbf{Know-how and skills.}
Skill libraries keep what an agent learned as executable code.
Voyager builds ``an ever-growing skill library of executable code for storing and retrieving complex behaviors'' in a game \cite{wang2024voyager}, and LRLL grows a robot skill library from experience \cite{tziafas2024lrll}.
These libraries only store know-how in a weak executability form, while knowledge as code is fully executable and verifiable.

\textbf{Exploratory programming.}
Human programmers learned by execution long before agents.
Exploratory and live programming study how programmers write code to find out what code does \cite{rein2019exploratory}.
This literature treats exploration as a programming practice, a way to write better programs.
Our account treats the same activity as a way to produce knowledge, where an exploratory program is a probe, and what it reveals becomes knowledge when it is written into persisted code.

\subsection{Epistemology}
\textbf{Knowing how (Ryle).}
Ryle distinguishes knowing that, which is knowledge of facts, from knowing how, which is the ability to do something, and holds that knowing how cannot be defined ``in terms of `knowing that'\,'' \cite[p.~32]{ryle1949concept}.
Code holds both: a constant or the expected value of a test is knowing that, and an algorithm is knowing how.
Code is the first medium where knowing how is written down and executed by the machine that reads it.

\textbf{The knowledge level (Newell).}
Newell holds that ``knowledge is to be characterized entirely functionally, in terms of what it does, not structurally'', and that ``knowledge of the world cannot be captured in a finite structure'' \cite[p.~10]{newell1981knowledge}.
For Newell, a data structure only represents knowledge, which belongs to the agent whose behavior it explains.
Code answers both objections, since it is a structure that does something.
Its text is the structure, its execution is what it does, and the executable knowledge lifecycle extends the structure each time a probe reveals something new about the world.

\textbf{Intervening (Hacking).}
Hacking moves philosophy of science from representing the world to intervening in it, since ``experimentation has a life of its own'' \cite[p.~xiii]{hacking1983representing}.
Code both represents the world and intervenes in it, since a program that is read states how the world is, and a program that is executed acts on the world and returns what it does.
Code is the first medium where the same text does both, since the program that states a belief about the world is also the experiment that tests it.

\subsection{Philosophy of programming}
\textbf{Programming as theory building.}
Naur argues that the theory of a program lives in its programmers, and that ``the program text and its documentation has proved insufficient as a carrier of some of the most important design ideas'' \cite{naur1985programming}.
We claim the opposite.
Coding agents keep nothing in mind from one session to the next, so the theory of a program must live outside their minds, and the code is the main place where it does.
The part of the theory that is in code is also the part that can be inspected, executed, verified and reproduced, which a theory held in the head of a programmer cannot be.

\section{Conclusion}
\label{sec:conclusion}

Code is the first knowledge medium that has all thirteen epistemic properties described in this paper.
Code is the only medium in which knowledge can act, be checked by a machine, be checked again later, and produce new knowledge.
Language carried knowledge for millennia, and it remains the medium in which we argue for what we know.
For millennia, we wrote knowledge to be read; we now write it to be run.

\printbibliography

\ifreview
\section*{Statements and Declarations}

\paragraph{Funding}
No funding was received for conducting this study.

\paragraph{Competing interests}
The author has no relevant financial or non-financial interests to disclose.

\paragraph{Ethics approval}
Not applicable. The corpus consists of the trajectories of the coding agents of the author, and no human participant was involved.

\paragraph{Data availability}
The instruments and their raw results are available for replication.
The trajectories contain private data and are not released.

\paragraph{Use of generative AI}
Generative AI has been used to prepare this paper.

\fi

\end{document}